\documentclass[aps,prd,twocolumn,showpacs]{revtex4} 

\usepackage{graphicx}
\usepackage{multirow}

\font\FermiSmallfont=cmssq8 scaled 1200

\def\LANLppthead#1{
\null 
\begin{center}\vskip -1.0truein{\hbox to 7.5truein {
\hfill
\vbox to 1in {\vfill \FermiSmallfont
              \hbox{#1}
              \vfill}
}}\vskip-0.0truein\end{center}}

\begin{document}

\title{Nuclear weak interaction rates in primordial nucleosynthesis}
\author{George M.\ Fuller$^1$ and Christel J.\ Smith$^2$}
\affiliation{$^1$Department of Physics, University of California, San Diego, La Jolla, CA
92093-0319}
\affiliation{$^2$Physics Department, Arizona State University, Tempe AZ, 85287-1504}

\date{\today}

\begin{abstract}

We calculate the weak interaction rates of selected light nuclei during the epoch of Big Bang Nucleosynthesis (BBN), and we assess the impact of these rates on nuclear abundance flow histories and on final light element abundance yields. We consider electron and electron antineutrino captures on $^3{\rm He}$ and $^7{\rm Be}$, and the reverse processes of positron capture and electron neutrino capture on $^3{\rm H}$ and $^7{\rm Li}$. We also compute the rates of positron and electron neutrino capture on $^6{\rm He}$. We calculate beta and positron decay transitions where appropriate. As expected, the final standard BBN abundance yields are little affected by addition of these weak processes, though there can be slight alterations of nuclear flow histories. However, non-standard BBN scenarios, {\it e.g.,} those involving out of equilibrium particle decay with energetic final state neutrinos, may be affected by these processes.  

\end{abstract}
\maketitle

\section{Introduction}

In this paper we examine a heretofore neglected side-story in primordial nucleosynthesis: the weak interaction rates of light nuclei. The fact that the abundances of these nuclei are small compared to those of the free protons and (sometimes) neutrons constitutes a persuasive reason for neglecting nuclear weak processes, at least when calculating nuclear abundance yields in standard Big Bang Nucleosynthesis (BBN). Indeed, the weak processes whose rates we calculate here are found to produce only very slight alterations in abundance yields and histories in standard BBN. 

Nevertheless, the physics of $e^\pm$ and $\nu_e/\bar\nu_e$ capture and beta/positron decay of light nuclei in BBN is interesting in itself. Elucidating this physics deepens our understanding of reaction flows in BBN. Moreover, nuclear weak processes may be more important in non-standard BBN models, especially those invoking decaying massive particles \cite{fks, afp, Petraki:2007gq, Kusenko:2004qc, Dolgov:2000ew, Fuller:2009zz, Cyburt:2010vz}. Recently models \cite{pastor} along these lines have been advanced as plausible variants to standard BBN.  

The calculation of the light element abundances as functions of the baryon and lepton numbers in the early universe and the comparison of these to observationally-derived abundances to determine key cosmological parameters is one of the great success stories of nuclear and particle physics and cosmology \cite{wag69, wfh, JYang, Walker:1991ap, 1990ApJ...358...47K, skm, kawano, kawano1, Nollett:2000fh,Schramm:1997vs,sfs,coulfac,Olive,Steigman1,steig, Steigman:2010zz, esposito}. The culmination of this enterprise was the determination of the baryon content of the universe from the observed deuterium abundance in high redshift QSO absorption systems \cite{Tytler, Omeara}, and the subsequent confirmation of this in measurements of the Cosmic Microwave Background (CMB) anisotropies \cite{WMAP1, 3yrwmap,Komatsu:2010fb}. Future CMB observations promise even higher precision, with Planck closing in on sub-one-percent uncertainty in the baryon-to-photon ratio \cite{bond,planck}.

As the scope and precision in cosmological observations have increased, puzzling issues in the standard BBN picture have emerged. The CMB-determined baryon-to-photon ratio and standard BBN predict a $^7{\rm Li}$ abundance a factor of two to three larger than that observed on the Spite plateau in hot, old halo stars \cite{spite, melendez}. Worse, recent claims of detection of isotope-shifted lithium absorption lines in a subset of these stars point to a $^6{\rm Li}$ abundance some three orders of magnitude larger than that expected in standard BBN \cite{asplund}. 

Neither of these problems is fatal for the standard model in our view. The first could be explained by {\it in situ} destruction of lithium in these stars via rotationally-driven turbulent diffusion or other mixing/diffusive processes \cite{ Steigman:2007xt, Korn:2006tv}. The second problem may not exist, as there are dissenting views on the interpretation of the stellar spectra \cite{2007A&A...473L..37C}. However, neither of these \lq\lq explanations\rq\rq\ is compelling either. Much is riding on the resolution of these questions, including possible insight into beyond-standard model massive particles and dark matter \cite{Cyburt:2006uv, Jedamzik:1999di}.

Perhaps the march to higher precision on the observational side should be matched by a sharpening in our understanding of BBN. In this spirit, the weak interaction processes involving light nuclei constitute one aspect of unexplored BBN physics that we {\it can} address. Very unlike the nuclear weak rate problem in stellar collapse \cite{FFNII, FFNIII, FFNI,FFNIV, fuller, Hix:2003fg, Langanke:2003ii, Sampaio:2002pe, Langanke:2002ab}, most of the nuclear data required for calculating the relevant rates in BBN have existed for a long time. Neutrino captures in some of the nuclear species considered here have been considered  for the post-explosion supernova environment \cite{Arcones:2008kv}.  In what follows we will discuss charged current weak interaction rates for the free nucleons and for $^3{\rm H}$, $^3{\rm He}$, $^6{\rm He}$, $^6{\rm Li}$, $^7{\rm Be}$, and $^7{\rm Li}$. We discuss the overall framework for lepton capture process and the relevant nuclear physics in section II. Results and discussion are given in section III, and conclusions in section IV.

\section{The Weak Interaction and Nucleosynthesis}

For weak interactions and nucleosynthesis in the early universe the salient feature of the universe we live in is its high (on a nuclear physics scale) entropy-per-baryon. In units of Boltzmann's constant per baryon this is $s/k_{\rm b} \approx 5.9\times{10}^{9}$, as calculated from the baryon-to-photon ratio $\eta \approx 6.11\times{10}^{-10}$ inferred at the epoch of photon decoupling by WMAP \cite{Komatsu:2010fb}. The large disorder implied by this entropy and the large number of photons per baryon has two consequences for our purposes: (1) In the BBN epoch the universe is radiation-dominated with number densities of electron/positron pairs and neutrino/antineutrino pairs comparable to that of photons and scaling with temperature as $T^3$, even down to temperatures {\it well below} the electron rest mass; and (2) BBN is essentially a freeze-out from nuclear statistical equilibrium (NSE) at high entropy. 

The weak interaction plays a key role in shaping the evolution of the early universe, especially as regards BBN. This evolution, like most of the interesting events in the very early universe, is a series of freeze-outs from equilibrium. First, both charged and neutral current neutrino scattering reactions on relativistic targets become slow compared to the expansion rate of the universe. This is Weak Decoupling. It occurs at temperatures $T \sim 1\,{\rm MeV}$. Though this decoupling is not sharp in time/temperature, eventually these neutrinos, decoupled into flavor states, simply free fall through spacetime, preserving their self-similar distribution in momentum space, with individual neutrino/antineutrino momenta redshifting like inverse scale factor \cite{Fuller:2008nt, Dodelson:2009ze, Giraud:2009tn,Bell:1998ds, dolgov, Ahluwalia:1997tr}. Given the high entropy and consequently large number of relativistic leptons per baryon, the rate of isospin flip engendered by the charged current reactions 
\begin{eqnarray}
& \nu_e+n\rightleftharpoons p+e^-, \label{nuen} \\
& \bar\nu_e+p\rightleftharpoons n+e^+, \label{nuebarp} \\
& n \rightleftharpoons p+e^-+\bar\nu_e. \label{ndecay}
\end{eqnarray}
for a free nucleon can be fast compared to the expansion rate. This maintains chemical equilibrium. However, since the rates of these processes scale as $T^5$, eventually this isospin flip rate will drop below the expansion rate, which goes as $T^2$. This is weak freeze-out. It is sometimes said that this occurs at $T\sim 0.7\,{\rm MeV}$, but in fact weak freeze-out also is not sharp in time/temperature. The neutron-to-proton ratio will continue to decrease slowly until neutrons are incorporated into alpha particles at $T\sim 0.1\,{\rm MeV}$, {\it i.e.,} NSE freeze-out. Only at this point will the abundances of light nuclei become appreciable.  

\begin{figure}

\includegraphics[trim=0in 1.5in 0in 1.5in, clip=true, width=3.5in]{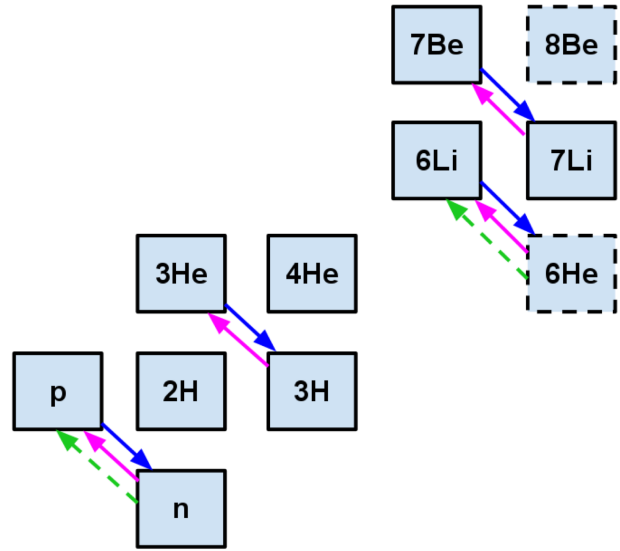}
\caption{The weak interaction transitions among light nuclei in BBN treated here. Beta decay processes are shown as dotted (green) arrows. Electron neutrino and positron capture reactions are shown as light (purple) arrows, while electron and electron antineutrino capture transitions are shown as darker (blue) arrows.}
\label{etafig1}
\end{figure}

The fastest and most favorable charged current weak transitions among the light nuclei in BBN are shown in Fig.~\ref{etafig1}. Of these, by far the most important are, of course, those proceeding on and through the free nucleons, {\it i.e.,} the reactions in Eq.~\ref{nuen}, Eq.~\ref{nuebarp}, and Eq.~\ref{ndecay}. In terms of raw leverage on the neutron-to-proton ratio and BBN abundance yields, the other weak processes shown in Fig.~\ref{etafig1} are not very significant, either because they are much slower than those occurring on the free nucleons or because the target nucleus abundances are so small, or both. 

\begin{figure}
\includegraphics[width=3.5in]{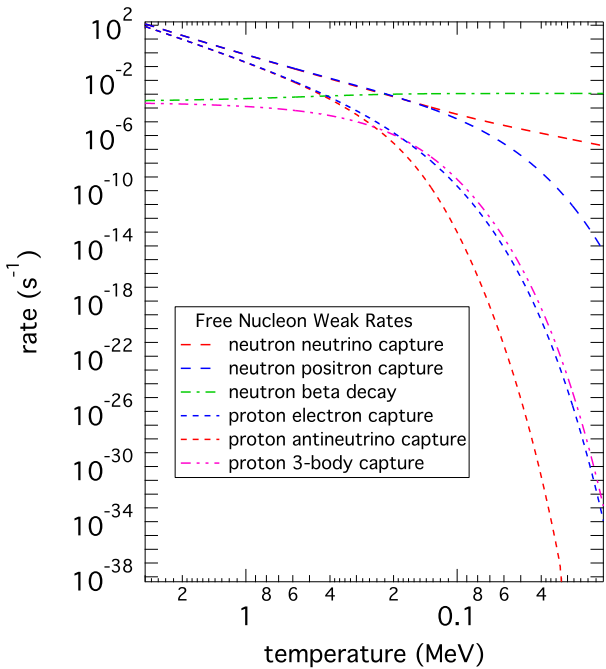}
\caption{The weak interaction rates (in ${\rm s}^{-1}$) for the free nucleons as functions of photon (plasma) temperature in MeV. The curves for the rates of the forward and reverse processes in Eq.~\ref{nuen}, Eq.~\ref{nuebarp}, and Eq.~\ref{ndecay} are as indicated in the legend.}
\label{etafigfree}
\end{figure}

The free nucleon charged current weak interaction rates and their effects are discussed in detail in Ref.~\cite{coulfac, sfs}. The rates for both the forward and reverse processes in Eq.~\ref{nuen}, Eq.~\ref{nuebarp}, and Eq.~\ref{ndecay} are shown in Fig.~\ref{etafigfree} for temperatures encompassing the weak freeze-out and BBN epochs. These rates are calculated in the same way as the light nuclear rates are calculated here: using an early universe code which consistently follows all thermodynamics, the Hubble expansion rate, and gives consistent photon and decoupled neutrino temperatures, and where the neutrino chemical potentials and degeneracy parameters are taken as zero. A version of the standard Kawano/Wagoner-Fowler-Hoyle code was modified as described in detail in Ref.~\cite{sfs} and Ref.~\cite{coulfac} to include independent weak interactions processes and neutrino and antineutrino energy distribution functions. This code was used to compute the effects of the rates discussed in this paper.

In the processes discussed below the total weak interaction rate for a given nucleus is a sum over parent states $i$ and daughter states $j$,
\begin{equation}
\lambda=\sum_i P_i \sum_j \lambda_{ij},
\end{equation}
where $P_i=(2J_i+1)e^{-E_i/T}/Z$ is the population factor at temperature $T$ for parent state $i$ with excitation energy $E_i$ and spin $J_i$.  The nuclear partition function is $Z$.  Here 
\begin{equation}
\lambda_{ij}= {{\ln 2}\over{ft_{ij}}}f_{ij}
\end{equation}
 is the rate connecting parent states $i$ to daughter state $j$, where $f_{ij}$ is the appropriate phase space factor and $ft_{ij}$ the corresponding $ft$-value.

\subsection{$^7{\rm Be}\rightleftharpoons{^7{\rm Li}}$}

Note that $^7{\rm Be}$ and $^7{\rm Li}$ are mirror nuclei. Each state in one nucleus is the isobaric analog of the corresponding state in the other. This isospin symmetry allows the measured nuclear weak transition data in this system to be leveraged to obtain weak matrix elements in some unmeasured branches. In the laboratory $^7{\rm Be}$ decays to $^7{\rm Li}$ by $K$-shell electron capture. However, in the early universe these species are ionized until the temperature falls to a few ${\rm eV}$, whereafter there will be a bound atomic electron and $K$-shell capture can occur. Consequently, in the early universe, during the BBN epoch, the weak transitions out of $^7{\rm Be}$ are through continuum electron capture, $\bar\nu_e$ capture, and by positron decay (and $e^-$ \& $\bar\nu_e$ capture) through a thermally-populated excited state,
\begin{eqnarray}
& \bar\nu_e + {^7{\rm Be}} \rightleftharpoons {^7{\rm Li}} + e^+, \label{barnue7} \\
& e^-+{^7{\rm Be}}\rightleftharpoons {^7{\rm Li}}+\nu_e,\label{ecap7} \\
&{^7{\rm Be}^\ast}\rightarrow {^7{\rm Li}}+e^++\nu_e,\label{pos}
\end{eqnarray}
where we also show the reverse processes of $\nu_e$ and $e^+$ capture on $^7{\rm Li}$.

For the $^7{\rm Be}$-destroying forward processes in Eq.~\ref{barnue7} and Eq.~\ref{ecap7} the positive ground-state-to-ground-state nuclear $Q$-value (nuclear mass of parent minus nuclear mass of daughter) is $Q_n \approx 0.3509\,{\rm MeV}$. The $ft$-value for this transition is measured to be $\log_{10} ft \approx 3.3$. The corresponding ground-state-to-first-excited-state (in $^7{\rm Li}$) transition has $Q_n \approx -0.1267\,{\rm MeV}$ and measured $\log_{10} ft \approx 3.5$. 

The threshold energy for the $\bar\nu_e$ in the reaction in Eq.~\ref{barnue7} is $E^\nu_{\rm th} \approx 0.1601\,{\rm MeV}$ for the ground-to-ground transition, and $E^\nu_{\rm th} \approx 0.6377\,{\rm MeV}$ for the ground-to-first transition. The electron threshold energy (including rest mass) in the Eq.~\ref{ecap7} ground-to-ground electron capture transition is $E^e_{\rm th} = m_e c^2\approx 0.511\,{\rm MeV}$, the electron rest mass, which corresponds to minimum final state $\bar\nu_e$ energy $E^\nu_{\rm th} =m_e c^2+Q_n$, and similarly for the ground-to-first transition.

The $\bar\nu_e$ capture rate per $^7{\rm Be}$ target nucleus for either of these transitions is
\begin{widetext}
\begin{equation}
\lambda_{\bar\nu_e} = {{\ln{2}}\over{ft}}\, \langle G\rangle\, b^5\, {{\left( {{ T }\over{ m_e c^2}} \right)}^5}
 \int_u^\infty{ x^2{\left( x+q \right)}^2{\left({1}\over{ e^{x+\eta_{\nu_e}}+1 }\right)}{\left(1-{{1 }\over{ e^{b\left(x+q\right)+\eta_e}+1 }}   \right)}  dx},
\label{PS1}
\end{equation}
while the electron capture rate for either of these transitions is given by
\begin{equation}
\lambda_{e^-} = {{\ln{2}}\over{ft}}\,\langle G\rangle\,{{\left( {{ T }\over{ m_e c^2}} \right)}^5}
 \int_u^\infty{ x^2{\left( x-q \right)}^2{\left({1}\over{ e^{x-q-\eta_e}+1 }\right)}{\left(1-{{1 }\over{ e^{x/b-\eta_{\nu_e}}+1 }}   \right)}  dx}.
\label{PS2}
\end{equation}
\end{widetext}
In these equations the ratio of neutrino temperature to plasma temperature is $b=T_\nu/T$, the $\nu_e$ degeneracy parameter (ratio of chemical potential to temperature) is $\eta_{\nu_e}$, while that for electrons is $\eta_e$, and $q=Q_n/T_\nu$ in Eq.~\ref{PS1}, but $q=Q_n/T$ in Eq.~\ref{PS2}. In both equations the lower limit on the integrals is the appropriate {\it neutrino} \lq\lq threshold\rq\rq\ energy scaled by temperature: $u\equiv E^\nu_{th}/T_\nu$ in Eq.~\ref{PS1}, where $E_{\rm th}^\nu$ is a true entrance channel $\bar\nu_e$ threshold energy; and $u\equiv E^\nu_{\rm th}/T$ in Eq.~\ref{PS2}, where $E^\nu_{\rm th} =m_e c^2+Q_n$ is the minimum final state neutrino energy.

In both Eq.~\ref{PS1} and Eq.~\ref{PS2} $\langle G\rangle$ is the average Coulomb wave correction factor, defined in Ref.~\cite{FFNI} and discussed in detail in Ref.~\cite{coulfac}. It is the average over the range of the integral of the product of the Fermi function $F(Z,w)$ and $w$, the ratio of electron momentum and energy. In the conditions characteristic of the BBN epoch, and for the nuclear transitions considered here, $\langle G\rangle \sim 1$. Here we use $\langle G\rangle =1$ for the forward process in Eq.~\ref{barnue7} and $\langle G\rangle =2$ for the forward rate in Eq.~\ref{ecap7}.  We performed a calculation of the electron capture process in Eq.~\ref{ecap7} with the full relativistic Coulomb correction as described in Ref.~\cite{coulfac} and find that on average $\langle G\rangle \sim 1.1$ over the relevant energy ranges.  Although we used $\langle G\rangle =2$ for this reaction here, we find that this difference has zero impact on the final abundance yields in standard BBN.  

The first excited state in $^7{\rm Be}$ is at $0.4292\,{\rm MeV}$ excitation energy, seemingly close to the $\sim 0.1\,{\rm MeV}$ temperature where this nuclear species is produced in BBN. Using mirror symmetry, and correcting for spin factors, we can estimate that the $ft$-value for the weak branch between this $J^\pi = {{1}\over{2}}^-$ state and the $^7{\rm Li}$\ $J^\pi = {{3}\over{2}}^-$ground state is $\log_{10} ft \approx 3.2$, with $Q_n \approx 0.7801\,{\rm MeV}$. Since this $Q_n$ value is bigger than the electron rest mass, the $\bar\nu_e$ threshold is $E_{\rm th}^\nu =0$. This large $Q$-value also implies that positron decay can proceed through this excited state, with the same $ft$-value. Once in this excited state there are allowed $\bar\nu_e$ and $e^-$ capture transitions to the $^7{\rm Li}$ first excited state, though these have much less favorable $Q$-values. The population factor for this state is $P_1 \approx 0.5\,\exp(-0.4292/T)$, where we approximate the partition function as the ground state spin degeneracy $2 J +1=4$.
In general, the population factor for this first excited state is only $\sim 1\%$ at $T\sim 0.1\,{\rm MeV}$, so all of these transitions contribute very little to the overall $^7{\rm Be}$ weak destruction rate in the regime where this species is being produced.

\begin{figure}
\includegraphics[width=3.5in]{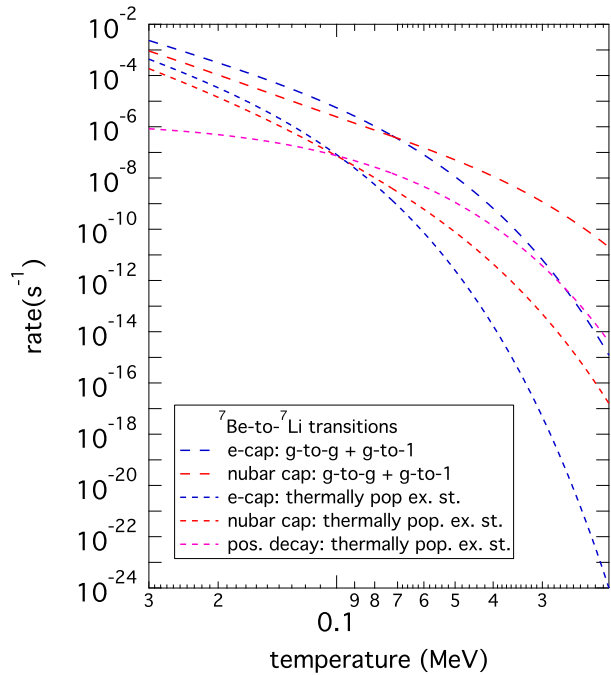}
\caption{Weak interaction rates for selected transitions in $^7{\rm Be}$ are given as a functions of temperature (MeV) in the early universe. For transitions from the ground state of $^7{\rm Be}$ to the $^7{\rm Li}$ ground and first excited states, the neutrino and electron capture rates are given by the light dashed (red) and darker dashed (blue) lines, respectively. The first excited state of $^7{\rm Be}$ can be thermally populated, and neutrino capture and electron capture rates for these transitions are given by the dot-dashed light (red) and dark (blue) lines, respectively. This thermally-populated state can also suffer positron decay and the rate for this is given by the very light (violet) dot-dashed line.}
\label{etafigx}
\end{figure}

The weak interaction rates for all of these transitions are shown as functions of temperature in Fig.~\ref{etafigx}. At temperatures in BBN where the abundance of $^7{\rm Be}$ comes up, $T\sim 0.1\,{\rm MeV}$, the dominant contribution to the total weak destruction rate of this species comes from neutrino and electron capture through the ground-to-ground and ground-to-first transitions. In fact, at this epoch the rates for neutrino capture and electron capture through these states are comparable. However, neutrino capture dominates over electron capture at low temperature, because the $e^\pm$-pair density dives for $T<80\,{\rm keV}$. This is despite the fact that the $e^\pm$ pair disappearance heats the photons relative to the neutrinos.

The rates for the processes proceeding through the thermally-populated first excited state in $^7{\rm Be}$ are also shown in Fig.~\ref{etafigx}. These are generally small. All three of neutrino capture, electron capture, and positron decay through/from this state are comparable at $T\sim 0.1\,{\rm MeV}$, and summed these rates comprise a few percent of the total weak destruction rate. Interestingly, at lower temperature positron decay through the thermally-populated first excited state of $^7{\rm Be}$, despite a tiny population factor, nevertheless dominates the electron capture rate on the ground state.

The rates of the reverse processes of $\nu_e$ and $e^+$ capture on $^7{\rm Li}$ are generally much slower than the forward rates of the processes in Eq.~\ref{barnue7} and Eq.~\ref{ecap7}. The rates of the processes proceeding through $^7{\rm Li}$ are $\sim 2\--3\%$ of the $^7{\rm Be}$ rates at $T\sim 0.1\,{\rm MeV}$. This is a result of the unfavorable $Q$-values in the $^7{\rm Li}\rightarrow{^7{\rm Be}}$ transition direction. The ground-to-ground $Q$-value is $Q_n \approx -0.3509\,{\rm MeV}$, while the ground-to-first transition has $Q_n\approx -0.7801\,{\rm MeV}$, and these branches have $\log_{10} ft= 3.3$ and $\log_{10} = 3.5$, respectively. The $\nu_e$ threshold energies for these transitions are $E_{\rm th}^\nu \approx 0.8619\,{\rm MeV}$ and $E_{\rm th}^\nu =0$, respectively.  

Thermal excitation of the first excited state in $^7{\rm Li}$, at excitation energy $0.4776\,{\rm MeV}$, produces a positive $Q$-value transition to the ground state of $^7{\rm Be}$ with $Q_n \approx 0.1267\,{\rm MeV}$, and this gives $E_{\rm th}^\nu \approx 0.3843\,{\rm MeV}$ for the $\nu_e$ capture channel, and minimum $\bar\nu_e$ energy $E_{\rm th}^\nu \approx 0.6377\,{\rm MeV}$ in the $e^+$ capture channel. By mirror symmetry, and correcting for spin differences, we can estimate that for this transition $\log_{10}ft\approx 3.2$. Again, the small population of the $^7{\rm Li}$ first excited state in the temperature regime where this species is principally produced causes the contribution of this transition to the overall rate to be negligible.

The $\nu_e$ capture rate per $^7{\rm Li}$ target nucleus for any of these transitions is
\begin{widetext}
\begin{equation}
\lambda_{\nu_e} = {{\ln{2}}\over{ft}}\, \langle G\rangle\, b^5\, {{\left( {{ T }\over{ m_e c^2}} \right)}^5}
 \int_u^\infty{ x^2{\left( x+q \right)}^2{\left({1}\over{ e^{x-\eta_{\nu_e}}+1 }\right)}{\left(1-{{1 }\over{ e^{b\left(x+q\right)-\eta_e}+1 }}   \right)}  dx},
\label{PS3}
\end{equation}
while the positron capture rate for these transitions can be calculated with
\begin{equation}
\lambda_{e^+} = {{\ln{2}}\over{ft}}\,\langle G\rangle\,{{\left( {{ T }\over{ m_e c^2}} \right)}^5}
 \int_u^\infty{ x^2{\left( x-q \right)}^2{\left({1}\over{ e^{x-q+\eta_e}+1 }\right)}{\left(1-{{1 }\over{ e^{x/b+\eta_{\nu_e}}+1 }}   \right)}  dx},
\label{PS4}
\end{equation}
where all notation is the same as in Eq.~\ref{PS1} and Eq.~\ref{PS2}. Here we take $\langle G\rangle =1$ for both the reverse process in Eq.~\ref{barnue7} and for the reverse process in Eq.~\ref{ecap7}.
\end{widetext}

\subsection{$^3{\rm He}\rightleftharpoons{^3{\rm H}}$}

The nuclei $^3{\rm He}$ and $^3{\rm H}$ are again mirrors, but this case is simpler than the beryllium-lithium system, in part because only the two $J^\pi = {{1}/{2}}^-$ ground states come into play. There are no excited states. The relevant weak interaction processes are
\begin{eqnarray}
& \bar\nu_e + {^3{\rm He}} \rightleftharpoons {^3{\rm H}} + e^+, \label{barnue3} \\
& e^-+{^3{\rm He}}\rightleftharpoons {^3{\rm H}}+\nu_e, \label{ecap3}\\
& {^3{\rm H}} \rightleftharpoons {^3{\rm He}}+e^-+\bar\nu_e. \label{decay3}
\end{eqnarray}
The forward process of $\bar\nu_e$ capture in Eq.~\ref{barnue3} has $Q_n \approx -0.5296\,{\rm MeV}$, giving a $\bar\nu_e$ energy threshold $E_{\rm th}^\nu \approx 1.0406\,{\rm MeV}$. This transition has measured $\log_{10}ft=3.1$. The rate can be estimated with Eq.~\ref{PS1} and using $\langle G\rangle =1$. The electron capture threshold, expressed as a minimum final state $\nu_e$ energy, is $E_{\rm th}^\nu =0$. This rate can be estimated with Eq.~\ref{PS2}, also with $\langle G\rangle =1$. 

The reverse processes of $\nu_e$ and $e^+$ capture in Eq.~\ref{barnue3} and Eq.~\ref{ecap3}, respectively, have more favorable $Q$-values. In this transition direction, we have $Q_n\approx 0.5296\,{\rm MeV}$, implying a $\nu_e$ threshold energy $E_{\rm th}^\nu =0$, and in the positron capture channel a minimum final state $\bar\nu_e$ energy $E_{\rm th}^\nu \approx 1.0406\,{\rm MeV}$. Both transitions have $\log_{10} ft=3.1$.

\begin{figure}
\includegraphics[width=3.5in]{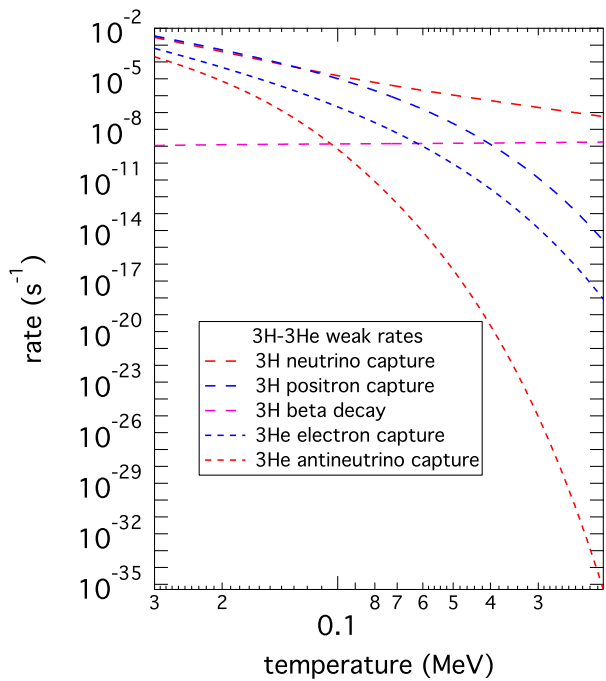}
\caption{Weak interaction rates for $^3{\rm H}$ and $^3{\rm He}$ are given as a functions of temperature (MeV) in the early universe. Curves are as labeled in the legend.}
\label{etafigy}
\end{figure}

In the laboratory, tritium decays via beta decay, the forward process in Eq.~\ref{decay3}, with a $12.33\,{\rm yr}$ half-life, corresponding to a rate $1.78\times{10}^{-9}\,{\rm s}^{-1}$. In relevant BBN conditions, where $T\sim 0.1\,{\rm MeV}$, this proves to be small compared to the lepton capture rates. The beta decay rate {\it rises} as the temperature of the universe decreases. This stems from easing of final state $\bar\nu_e$ and electron phase space blocking at lower temperature. Free neutron beta decay also shows this phenomenon, as is evident in Fig.~\ref{etafigfree}.
The rates for all of the $^3{\rm He}$ and $^3{\rm H}$ weak processes are shown in Fig.~\ref{etafigy}.

\subsection{$^6{\rm He}\rightleftharpoons{^6{\rm Li}}$}

The mass six system plays a very minor role in standard BBN but, as discussed above, is nevertheless a focus of recent interest because of the claimed detection of $^6{\rm Li}$ on the surfaces of hot, old halo stars. Only recently \cite{Boyd} has $^6{\rm He}$, along with the $^7{\rm Li}\left(^3{\rm H},\alpha\right){^6{\rm He}}$ reaction been incorporated into the BBN reaction network \-- all to very little effect. However, the mass six system is intriguing from a weak interaction standpoint.

The weak reactions of interest are
\begin{eqnarray}
& \nu_e + {^6{\rm He}} \rightleftharpoons {^6{\rm Li}} + e^-, \label{barnue6} \\
& e^++{^6{\rm He}}\rightleftharpoons {^6{\rm Li}}+\bar\nu_e, \label{ecap6} \\
& {^6{\rm He}} \rightleftharpoons {^6{\rm Li}}+e^-+\bar\nu_e. \label{decay6}
\end{eqnarray}
For the forward processes in these equations there is an impressive $Q_n\approx 4.0207\,{\rm MeV}$, and a respectably large weak matrix element, as implied by the measured $\log_{10} ft = 2.9$. The threshold $\nu_e$ energy in the forward process in Eq.~\ref{barnue6} is $E_{\rm th}^\nu=0$, while the minimum final state $\bar\nu_e$ energy in the forward process in Eq.~\ref{ecap6} is $E_{\rm th}^\nu\approx 4.5317\,{\rm MeV}$. The rates for these two channels can be estimated using Eq.~\ref{PS3} and Eq.~\ref{PS4}, respectively.

In the laboratory $^6{\rm He}$ decays via beta decay, Eq.~\ref{decay6}, with a half-life of $0.808\,{\rm s}$, implying an unblocked decay rate $\lambda_{beta} \approx 0.858\,{\rm s}^{-1}$. In the conditions relevant for BBN where this rate is dominant, there is negligible final state $e^-$ and $\bar\nu_e$ blocking.

The reverse processes in Eq.~\ref{barnue6}, Eq.~\ref{ecap6}, and Eq.~\ref{decay6} all have, of course,  a highly negative and unfavorable $Q_n$ value, rendering the rates in these reverse channels negligible at all relevant BBN temperatures. There are excited states for these nuclei. They are at  high enough excitation energies that their thermal populations are tiny, but these excitation energies are not high enough to counter the large negative $Q$-values.




%






\section{Results and Discussion}

\begin{table}
\caption{}
\begin{tabular}{ | l | c | }
\multicolumn{2}{c}{Weak Reactions Added to the BBN Code} \\
\hline
Reaction & Effect\\
\hline\hline
$\nu_e\ +\ ^3$H$\ \rightarrow\ ^3$He$\ +\ e^-$&\multirow{3}{*}{tiny}\\
$e^+\ +\ ^3$H$\ \rightarrow\ ^3$He$\ +\ \bar\nu_e$&\\
$^3$H$\ \rightarrow\ ^3$He$\ +\ e^- +\ \bar\nu_e$&\\

\hline
$\bar\nu_e\ +\ ^3$He$\ \rightarrow\ ^3$H$\ +\ e^+$&\multirow{2}{*}{none}\\
$e^-\ +\ ^3$He$\ \rightarrow\ ^3$H$\ +\ \nu_e$&\\

\hline
$\nu_e\ +\ ^7$Li$\ \rightarrow\ ^7$Be$\ +\ e^-$&\multirow{2}{*}{none}\\
$e^+\ +\ ^7$Li$\ \rightarrow\ ^7$Be$\ +\ \bar\nu_e$&\\
\hline
$\bar\nu_e\ +\ ^7$Be$\ \rightarrow\ ^7$Li$\ +\ e^+$&\multirow{2}{*}{none}\\
$e^-\ +\ ^7$Be$\ \rightarrow\ ^7$Li$\ +\ \nu_e$&\\

\hline
$\nu_e\ +\ ^6$He$\ \rightarrow\ ^6$Li$\ +\ e^-$&\multirow{3}{*}{small}\\
$e^+\ +\ ^6$He$\ \rightarrow\ ^6$Li$\ +\ \bar\nu_e$&\\
$^6$He$\ \rightarrow\  ^6$Li$\ +\ e^-  +\ \bar\nu_e$&\\

\hline

\hline
\end{tabular}
\end{table}

The weak interaction and, in particular, the comparison of the neutron-proton weak interconversion rates to the expansion rate of the universe, is the cornerstone of BBN. This is key contributor to the power BBN has to constrain speculative new physics in the particle sector. None of the nuclear weak processes discussed here change these results, either because the nuclear target abundances are so small, or because the rates of these are small at relevant epochs.

\begin{figure}
\includegraphics[width=3.1in, angle=270]{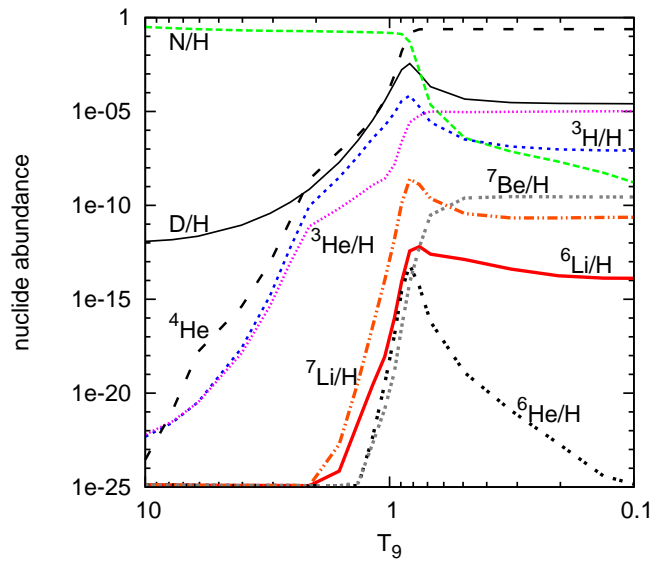}
\caption{Nuclear abundance as a function of temperature $T_9$ for key light nuclear species. Here $T_9=T/10^9\,{\rm K}$ and the corresponding temperature in MeV is $T\approx 0.8617\, T_9$. The deuterium ($^2{\rm H}$) abundance relative to hydrogen is labeled D/H and the free neutron abundance relative to hydrogen is labeled N/H. Other abundances are relative to hydrogen as labeled, except for $^4{\rm He}$ which is a mass fraction.  }
\label{etafig2}
\end{figure}

The nuclear abundances as a function of temperature during the BBN process are shown in Fig.~\ref{etafig2}. These abundances were calculated with a modified version of the standard BBN code derived from Wagoner, Fowler, and Hoyle \cite{wfh} and Kawano \cite{kawano,kawano1}, and described in detail in Smith and Fuller \cite{coulfac} and Smith, Fuller, and Smith \cite{sfs}. This calculation includes the rates of the weak interaction processes discussed in the last section. 

It is clear from Fig.~\ref{etafig2} that the abundances of the lithium and beryllium isotopes do not come up until the temperature falls to $T\sim 0.1\,{\rm MeV}$, and even then they remain very small. In fact, $^6{\rm He}$ contributes in the run-up towards the peak in the $^6{\rm Li}$ abundance, but in the end is responsible for only a tiny fraction of the ultimate $^6{\rm Li}$ yield, which stems mostly from ${^4{\rm He}}\left( {^2{\rm H}},\gamma \right){^6{\rm Li}}$ \cite{Nollett:1996ef}. As we will discuss below, the weak interaction  efficiently and quickly converts $^6{\rm He}$ to $^6{\rm Li}$.

The tritium and $^3{\rm He}$ abundances come up earlier, tracking the Saha equation NSE predictions until $T\approx 0.2\,{\rm MeV}$. Thereafter, as the temperature drops, the $^3{\rm He}$ abundance drops below that of $^3{\rm H}$, until both reach a peak near $T\approx 90\,{\rm keV}$. Subsequently, $^3{\rm H}$ is converted to $^3{\rm He}$ by weak lepton capture reactions and, below $T\sim 20\,{\rm keV}$, by tritium beta decay. 

Addition of these weak interaction processes to the BBN calculation produces scant change in final abundance yields.  Table I summarizes the yield alterations.  In this table, \lq\lq none\rq\rq\ means final abundance yield change of one part in $10^6$ or less, while \lq\lq tiny\rq\rq\ means changes of one part in $10^4$ or so.  For example, the weak processes with $^3$H as a target result in $10^{-2} \%$ decreases in deuterium and $^3$H final yields, a $10^{-2} \%$ increase in $^3$He, and a $10^{-3}\%$ increase in $^7$Li.  Adding in the weak rates for $^7$Be decreases the final abundances of $^7$Li, $^2$H, $^3$He, and $^4$He by about $10^{-4}\%$ to $10^{-5}\%$.

Although addition of the $^6$He weak rates makes a negligible alteration in the final $^6$Li abundance yield, this is only because $^6$He does not figure in the primary $^4$He($^2$H,$\gamma)^6$Li production reaction at later times.  However, as is evident from the $^6$He and $^6$Li yield curves in Fig.~\ref{etafig2}, near $T_9=1$ when $^6$Li is building up, $^6$He contributes to the $^6$Li abundance.  Since $^6$He is quickly converted to $^6$Li via weak interactions at this temperature, inclusion of these rates does alter the nuclear flow history slightly, albeit with no effect on the final abundance yield of $^6$Li.  Nevertheless, this alteration in history garners a \lq\lq small\rq\rq\ label in Table I.

\begin{figure}
\includegraphics[width=3.5in]{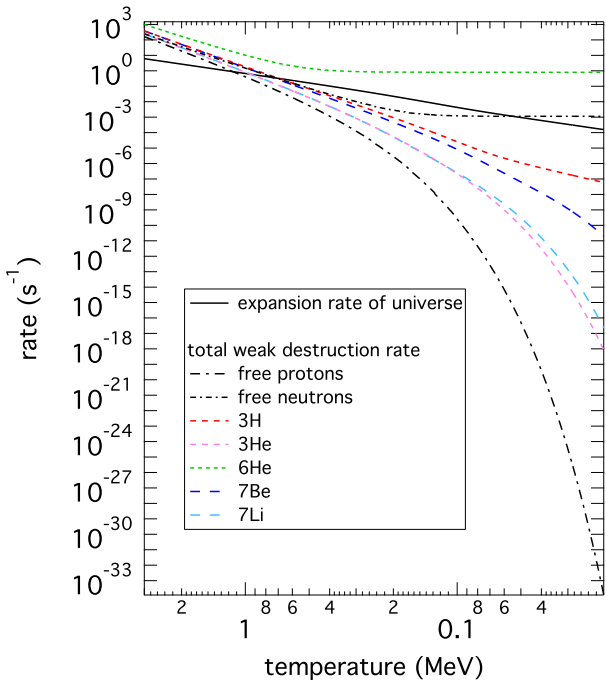}
\caption{Total weak destruction rates per target are shown as functions of temperature (in MeV) for various nuclear species, as labeled. Also shown is the expansion rate of the universe, given by the heavy (black) line.}
\label{etafig3}
\end{figure}

In Fig.~\ref{etafig3} we show selected weak interaction rates per target as a function of temperature during the weak freeze-out and BBN epochs. This figure also shows the expansion rate of the universe through these epochs. The overall weak destruction rates for target nuclei shown in this figure are calculated by summing the relevant transitions discussed in the last section. 

The general trends of the rates on this plot stem from the conditions in the early universe. As the universe expands and the temperature of both the plasma and the decoupled neutrinos drop, lepton capture rates fall dramatically. Two issues exacerbate this trend.

Although the neutrino temperature starts out the same as the plasma temperature ({\it i.e.,} for temperatures well above decoupling), as the universe expands this is no longer the case. As the temperature drops and electromagnetic equilibrium shifts to include fewer $e^\pm$ pairs, the entropy that these carried gets shifted to the photons. This means that the plasma temperature drops less steeply than that of the decoupled neutrinos. The neutrino temperature simply redshifts with inverse scale factor. By the time the pairs are gone ($T\le 20\,{\rm keV}$), the ratio of neutrino-to-plasma temperatures is ${\left( 4/11 \right)}^{1/3}$. Even before this, the neutrino temperature pulls away from the plasma temperature when $T\sim m_e c^2$. The upshot is that although electron or positron capture may be larger than $\bar\nu_e$ or $\nu_e$ capture early on, this will not necessarily be true later.

Cases in point are the free nucleons. The free neutron $e^+$ and $\nu_e$ capture rates are large to begin with, dominating over the beta decay rate, and larger than the proton $e^-$ and $\bar\nu_e$ capture rates. However, as the universe expands and cools, the neutron destruction rate asymptotes to the vacuum beta decay rate, while the proton lepton capture rates crash. This behavior is readily apparent in Fig.~\ref{etafig3}. Early on all of these rates are larger than the expansion rate. This is where weak neutron-proton equilibrium is maintained, and in this regime the neutron-to-proton ratio falls slowly with decreasing temperature in accord with the Saha equation. Once these rates fall below the expansion rate (weak freeze-out), the neutron-to-proton ratio will fall a little, mostly as a result of free neutron beta decay. Of course, once essentially all neutrons are locked up in alpha particles at $T\sim 0.1\,{\rm MeV}$, the fact that the neutron beta decay rate again becomes larger than the expansion rate at $T <45\,{\rm keV}$ is irrelevant.

Interestingly, for any temperature, putting a neutron into $^6{\rm He}$ causes it to be converted to a proton much faster than if it were a free particle. Moreover, this conversion rate is {\it always} faster than the expansion rate of the universe. This, of course, has negligible effect on the neutron-to-proton ratio at any time in BBN. At temperatures $T> 0.5\,{\rm MeV}$, where lepton captures make $^6{\rm He}\rightarrow {^6{\rm Li}}$ very fast, there is no $^6{\rm He}$, as Fig.~\ref{etafig2} makes clear. At lower temperatures, say $T=0.09\,{\rm MeV}$, near the $^6{\rm He}$ abundance peak, only about one neutron in ${10}^{8}$ resides in $^6{\rm He}$! 

This is a classic example of a phenomenon familiar in stellar collapse/supernovae: very neutron-rich nuclides have large weak strength and fast decay rates; yet may have small abundances. In BBN the conditions are neutron {\it deficient}, not neutron-rich, so the rarity of a neutron-rich nuclide like $^6{\rm He}$ is even more pronounced. What then about $^6{\rm Be} \rightarrow {^6{\rm Li}}$? This transition is the mirror of the $^6{\rm He}\rightarrow{^6{\rm Li}}$ ground-to-ground, $J^\pi = 0^+ \rightarrow 1^+$ transition, for which $\log_{10} ft=2.9$ in  both cases. It has $Q_n\approx 3.777\,{\rm MeV}$, respectable, but slightly smaller than in the mirror channel. However, $^6{\rm Be}$ is particle unstable, with a very fast strong interaction decay into an alpha particle and two protons. In general, the relatively low temperatures where BBN occurs, along with the consequently substantial Coulomb barriers prevent the assembly of proton-rich nuclides for which weak decay rates would be significant.

The $^3{\rm H}$ and $^3{\rm He}$ abundances come up relatively early, as discussed above, and in this case the lepton capture-mediated rates can dominate over tritium beta decay. Including these rates in the BBN calculation makes a very small change (one part in ${10}^4$) in the ultimate lithium and $^3{\rm He}$ abundances. In Fig.~\ref{etafig3} the total weak destruction rate of $^3{\rm H}$ shows a break to a shallower slope for temperatures $T<60\,{\rm keV}$. This is where the neutrino capture rate becomes larger than the positron capture rate. This latter rate dives as the $e^\pm$ pair density goes down exponentially with temperature at sufficiently low temperature.

Similar behavior is evident in Fig.~\ref{etafig3} for the $^7{\rm Be}$ weak destruction rate. A break in the negative slope in the rate at $T\sim 60\,{\rm keV}$ stems from the disappearance of $e^\pm$ pairs. In this case, however, the $Q_n$ is so small that there is a $\bar\nu_e$ energy threshold, which is significant given the temperature $T_\nu$ characteristic of the neutrinos at this epoch. Note that both the $^3{\rm He}$ and the $^7{\rm Be}$ weak destruction rates are orders of magnitude smaller than the expansion rate after these species are produced in BBN. 


\section{Conclusion}

The weak interactions (lepton capture and beta and positron decay) involving light nuclei in BBN have little effect on standard BBN abundance yields, either because these nuclei have small abundances or because the weak rates themselves are tiny, or both. Nevertheless, these weak transitions are an interesting story in themselves. In non-standard models, especially those involving decaying particles with high energy final state neutrinos, the rates of these processes may be fast enough to alter abundance yields significantly. 

We have identified key weak nuclear transitions in BBN and elucidated the nuclear physics underlying these, using measured data and exploiting mirror symmetry to find matrix elements for excited state transitions in some cases. We have also provided a simple prescription for calculating the relevant lepton capture rates for these transitions. The rates calculated here, along with this prescription, will be placed on Big Bang Online \cite{bigbangonline}.

The weak interaction, and the interconversion of neutrons and protons by it, is a foundational component of BBN. The weak processes involving free nucleons are by far the dominant arbiters of what happens in BBN. Ultimately, like so many aspects of BBN and the physics of the early universe, this is a function of the high entropy per baryon in the universe. High entropy immediately dictates that most nucleons cannot reside in nuclei, but rather must be free. Indeed, in broad brush, standard BBN puts nearly all neutrons into alpha particles at $T\sim 0.1\,{\rm MeV}$. For example, we have one in 16 nucleons in $^4{\rm He}$, but only one in ${10}^5$ in deuterons, and a scant one in ${10}^{10}$ in $^7{\rm Be}$/$^7{\rm Li}$.  

Although the Fermi and Gamow-Teller matrix elements are large for the free nucleons, they can be as big or even bigger for select transitions involving the light nuclei. Moreover, transitions in the light nuclei can sometimes have $Q$-value advantages over free nucleon weak processes. An example of both advantages can be found in the $^6{\rm He}\rightarrow{^6{\rm Li}}$ transition discussed in the last section. A neutron placed in a $^6{\rm He}$ nucleus during BBN is converted to a proton much faster than if it were a free neutron, and this process is always faster than the expansion rate of the universe. Of course, the mass fraction of $^6{\rm He}$ is tiny and so the leverage of this weak decay on BBN is negligible. The low abundance of this species is because the BBN temperatures are relatively low, Coulomb barriers are therefore non-negligible, the mass of $^6{\rm He}$ is relatively high, and conditions in BBN are neutron-poor. Ultimately, the overriding reason is because the entropy is high and BBN is nearly a freeze-out from equilibrium conditions.

This suggests that if weak interactions in the light nuclei are ever to be important in BBN, it will be in conditions where nuclear chemical equilibrium breaks down differently than in the standard picture, or where non-thermal, non-beta equilibrium distributions of neutrinos exist. An example of the latter scenario is where there are particles that have number densities comparable to photons at (or before) the BBN epoch and decay with an appreciable branching ratio into light neutrinos. 

A specific model along these lines involves sterile neutrinos with rest masses $\sim 100\,{\rm MeV}$ and with vacuum mixing with active neutrinos at the level of one part in $\sim {10}^8$. These particles would be in thermal and chemical equilibrium at temperature scales $T> 1\,{\rm GeV}$, but could have lifetimes against decay into three light neutrinos $> 1\,{\rm s}$, leading to energetic neutrinos with non-thermal energy spectra \cite{fks}. Since the cross sections for the neutrino capture processes discussed here are highly energy dependent, they could alter BBN abundance yields and, indeed, the whole BBN paradigm. In almost all cases drastic alterations of BBN along these lines leads to unacceptable light element abundances. This, in turn, is a way of constraining particle physics which may be inaccessible in the laboratory. The simple approximate rate integral expressions given in this paper are adequate for roughing out the effects of decaying particles and thereby carrying out this parameter space constraint procedure.

We know that the relic neutrino background must be there during weak freeze out and BBN. Were it not, and were there no neutrino and antineutrino captures on free nucleons, the BBN abundance yields would be in gross conflict with observation \cite{sfs}. However, we do not know whether this relic neutrino background is altered subsequent to the BBN epoch. There is as yet no detection of the neutrino rest mass from CMB and large scale structure observations. Much about the relic neutrino background and the neutrino sector in the early universe remains mysterious.

\vskip 1.0in

\begin{acknowledgments}
We would like to acknowledge helpful discussions with Richard Boyd, Wick Haxton, Chad Kishimoto, Lawrence Krauss, and Cecelia Lunardini.  This work was supported in part by NSF Grant No. PHY-06-53626 at UCSD and CJS would like to thank ASU for support.
\end{acknowledgments}

\bibliography{mybiblio}



\end{document}